\documentclass[12pt,amsmath,prb,preprint]{revtex4}

\usepackage{color}
\usepackage{amsfonts}
\usepackage{amssymb}
\usepackage{graphicx}

\begin{document}
\title{Molecular Heat Engines: Quantum Coherence Effects}
\author{Feng Chen}
\affiliation{Department of Physics, University of California San Diego, La Jolla, CA 92093, USA}
\author{Yi Gao}
\affiliation{Department of Chemistry \& Biochemistry, University of California San Diego, La Jolla, CA 92093, USA}
\altaffiliation{PIMCO, Investment Management, Newport Beach, CA 92660, USA}
\author{Michael Galperin}
\email{migalperin@ucsd.edu}
\affiliation{Department of Chemistry \& Biochemistry, University of California San Diego, La Jolla, CA 92093, USA}

\begin{abstract}
Recent developments in nanoscale experimental techniques made it possible to utilize
single molecule junctions as devices for electronics and energy transfer
with quantum coherence playing an important role in their thermoelectric characteristics.
Theoretical studies on the efficiency of nanoscale devices usually employ rate (Pauli) equations,
which do not account for quantum coherence. Therefore, the question whether quantum coherence
could improve the efficiency of a molecular device cannot be fully addressed within such considerations.
Here, we employ a nonequilibrium Green function approach to study the effects of quantum coherence 
and dephasing on the thermoelectric performance of molecular heat engines.
Within a generic bichromophoric donor-bridge-acceptor junction model, we show 
that quantum coherence may increase efficiency compared to quasi-classical 
(rate equation) predictions and that pure dephasing and dissipation destroy this effect.
\end{abstract}

\maketitle

%%%%%%%%%%%%%%%%%%%%%%%%%%%%%%%%%%%%%%%%%%
\section{Introduction}
Single molecules are used as building blocks of molecular devices for
electronics, biosensors, nanoscale motors, controllable chemical reactivity and
energy transfer~\cite{AviramRatnerCPL74,RatnerNatNano13,visionsNatNano13}.
The development of nano-fabrication led to tremendous progress in
the ability to detect and manipulate molecules on surfaces and in 
junctions~\cite{CenturionPRL12,HoJCP13,SchaffertNatMater13,DongHouNature13,HoScience14,LortscherNatNano16}.
Due to the small size of nanodevices, their characterization is
necessarily quantum, and interference is expected
to be crucial in the response of molecular electronic devices.
Experimental measurements have demonstrated quantum coherence effects
on the transport~\cite{MayorWeberAngChim03,VenkataramanHybertsenNatNano12,WeberPRL12} and optical response~\cite{VanDuyneNL12,ApkarianACSNano14} of molecular junctions.
The importance of quantum coherence in energy transfer was demonstrated
in experimental studies of the initial stages of photosynthesis~\cite{FlemingScience07,EngelFlemingNature07,EngelPNAS10,FlemingPCCP10}.

Understanding and controlling the combined motion of charges and excitations (energy)
is crucial for the development of new materials and state-of-the-art guiding principles
for building efficient energy conversion and storage devices.
Historically, this research has been focused on thermoelectric properties in bulk 
materials~\cite{Ioffe_1957}.
With the development of nanofabrication techniques, the study of
thermoelectric properties at the nanoscale attracted a lot of attention
experimentally~\cite{MajumdarMcEuenPRL01,MajumdarScience07,MajumdarCPL10,VenkataramanNL12,TsutsuiSciRep13,EsslingerScience13,ChabinycNatMater14,ReddyNatNano14,LeePRL14}
and theoretically~\cite{PaulssonDattaPRB03,LinkePRL05,MGNitzanRatnerMP08,LiuChenACSNano09,LiuChenPRB09,DubiDiVentraNL09,WegewijsFlensbergPRB10,FranssonMGPCCP11,EntinWohlmanImryAharonyPRB10,EntinWohlmanAharonyPRB12,ThygesenJCompEl12,SanchezLopezPRL13}.
The small size of these nanodevices gives rise to new physical phenomena
(such as quantum coherence) that are not present at the macroscopic level, and which promise
to improve the performance of energy conversion.
These studies are concerned with characterizing charge and
energy fluxes in the system. 

A closely related set of works utilized the thermodynamic 
approach to determine the efficiency of photoelectric devices~\cite{EspositoPRB09,EinaxDierlNitzanJPCC11,CreatorePRL13,EinaxNitzanJPCC14,EinaxNitzanARXIV15}.
With rate (Pauli) equations employed in the thermodynamic description  of such devices,
quantum coherence could not be fully taken into account (see below).
The effect of coherence on the thermodynamics of quantum heat engines consisting of
n-level systems coupled to thermal bath(s) was considered in a number of
publications~\cite{HarbolaMukamelPRA12,MukamelPNAS13,HarbolaPRA13,MustecapliogluPRA15,LeggioPRA15} at the Lindblad-Redfield level of theory.
These considerations were either restricted to closed systems or disregarded charge
transfer between the system and the baths (the latter is inherent in molecular devices).
Thus, naturally also in thermodynamic studies, the question of how quantum coherence 
affects a molecular device performance can be raised. 

We note in passing that the distinction between populations 
and coherences (diagonal and off-diagonal elements of the density matrix) is basis dependent.
For example, transforming the local basis to the eigenbasis of the system
accounts for coherences of the local basis simply by converting
them into populations in the eigenbasis. For an isolated system,
local basis coherences are taken into account exactly
as a result of such a transformation. In the presence of baths, 
for the consideration to be complete one has to account also for bath-induced coherences 
between the eigenstates of the system. The latter cannot be treated properly at 
the Lindblad-Redfield quantum master equation (QME) level of theory
(one has to go to at least the fourth order in system-bath coupling to account for the coherences~\cite{LeijnseWegewijsPRB08}), 
while approximate consideration may lead to qualitative failures (see, e.g.,~\cite{EspositoMGJPCC10,EspMGPRB15,GaoMGJCP16_2}).
In addition, Lindblad-Redfield QME does not allow to model the gradual transition from 
coherent to incoherent transport. The latter is the focus of our study.  

Here, we consider a generic donor-bridge-acceptor (DBA) molecular system, which is
coupled to fermionic baths and is driven against the bias applied by solar photons.
This setup is a simple model for a continuous steady-state heat engine~\cite{KosloffLevyAnnRevPhysChem14,VanDenBroeckEPL15,KosloffEntropy17},
whose thermoelectric efficiency was previously considered in~\cite{EspositoPRB09,EinaxDierlNitzanJPCC11,EinaxNitzanJPCC14}
with the effects of quantum coherence disregarded.
The latter were shown to play an important role in the charge and energy transport in
similar models of DBA molecular junctions~\cite{PeskinMGJCP12,WhitePeskinMGPRB13}.

We utilize nonequilibirum Green function (NEGF) methodology~\cite{HaugJauho_2008},
which is capable of accounting for quantum coherence in an open nonequilibrium system,
to study the effects of quantum coherence on the average efficiency of  photoelectric molecular devices.
For simplicity, our consideration is restricted to a non-interacting 
(electron-electron and electron-vibration interactions are  disregarded)
molecular system,
although intra-system interactions in principle can be taken into account within 
many-body flavors of the methodology~\cite{WernerRMP14,WhiteOchoaMGJPCC14,ChenOchoaMGJCP17,MGChemSocRev17,MiwaChenMG17}.
We show that quantum coherence may lead to an increase and a decrease in the efficiency of the device
and study the transition to a quasi-classical regime by destroying coherence with pure dephasing
or dissipation. The former  is achieved by employing a B{\" u}ttiker probe~\cite{DattaPRB07},
and the latter is induced by increasing the strength of the system-contacts coupling.
We note in passing that although a general formulation of quantum thermodynamics 
for current-carrying junctions has not yet been established~\cite{EspOchoaMGPRB15,NitzanPRB16},
the formulation is clear as long as the junction operates in a steady-state~\cite{GaspardNJP15,NessEntropy17}.
The latter is the situation considered~here.

The paper is structured as follows: In Section~\ref{model}, we introduce the model and discuss
the technical details of the simulations. 
We present the results of the numerical simulations and compare them with previously published 
(coherence-free) studies in Section~\ref{numres}. We draw conclusions in Section~\ref{conclude}.

\section{Model}\label{model}
\begin{figure}[htbp]
\centering\includegraphics[width=0.5\linewidth]{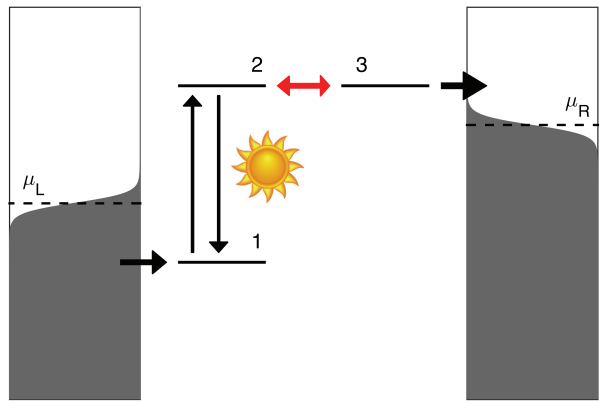}
\caption{\label{fig1}
Sketch of the molecular donor-bridge-acceptor (DBA) heat engine model.
}
\end{figure}

We consider a molecular junction comprised of a DBA molecular complex 
coupled to metallic contacts $L$ and $R$ (see Figure \ref{fig1}).
The contacts are equilibrium reservoirs of free charge carriers 
maintained at the same temperature $T$. The junction is biased so that the electrochemical potential
of contact $L$ is lower than that of $R$, $\mu_L < \mu_R$.
The donor and the acceptor are modeled as two-level 
(highest occupied molecular orbital-lowest unoccupied molecular orbital, HOMO-LUMO) systems.
Following \cite{EinaxNitzanJPCC14}, we assume that the HOMO of the acceptor is always populated
(i.e., does not participate in transport), and thus can be disregarded.
The bridge provides super-exchange coupling between the donor and the acceptor and is accounted 
for by effective electron hopping matrix element~$t$.
The donor is subjected to solar radiation which is modeled by coupling to
a thermal bath $S$ of high (solar) temperature $T_S$ ($T_S\gg T$).
Electron transfer is allowed between the LUMOs of the donor and the acceptor;
thus, solar radiation drives the electronic flux against the applied bias
(heat engine).
Bridge-induced coupling between the LUMOs (levels $2$ and $3$ in Figure \ref{fig1})
is the cause of intra-molecular  quantum coherence.
For an isolated two-level system, such coherence leads to permanent Rabi oscillations
in the electron population on both levels.
In our model, these oscillations are damped by coupling to baths (solar radiation and contacts) 
with the damping rate depending on the strength of the couplings.

%%%%%%%%%%%%%%%%%%%%%%%%%%%%%%%%%%%%%%%%%%%%%
%%%%%%%%%%%%%%%%%%%%%%%%%%%%%%%%%%%%%%%%%%%%%

The Hamiltonian of the model is
\begin{equation}
\label{H}
\hat H = \hat H_M + \sum_{B=L,R,S}(\hat H_B + \hat V_B),
\end{equation}
where the $\hat H_M$ and $\hat H_B$ Hamiltonians represent the molecule ($M$), 
contacts ($L$ and $R$) and thermal bath ($S$), respectively. 
$\hat V_B$ couples the molecule to the baths (contacts and Sun radiation). 
Explicit expressions {are}
\begin{equation}
\label{Hexp}
\begin{split}
 &\hat H_M = \sum_{m=1}^3\varepsilon_m\hat d_m^\dagger \hat d_m
 +
 t\left(\hat d_2^\dagger\hat d_3 + \hat d_3^\dagger\hat d_2\right)
 \\
 &\hat H_{L(R)} = \sum_{k\in L(R)} \varepsilon_k\hat c_k^\dagger\hat c_k; \qquad\quad
 \hat H_S = \sum_{\alpha\in S} \omega_\alpha \hat a^\dagger_\alpha \hat a_\alpha
 \\
&\hat V_{L} = \sum_{\ell\in L}\left( V_{\ell}\hat d^\dagger_1\hat c_\ell + H.c.\right); \quad
 \hat V_{R} = \sum_{r\in R}\left( V_{r}\hat d^\dagger_3\hat c_r + H.c.\right)
 \\
 &\hat V_S = \sum_{\alpha\in S}\left( U_\alpha \hat D^\dagger\,\hat a_\alpha + H.c.\right).
 \end{split}
\end{equation}

Here, $\hat d_m^\dagger$ ($\hat d_m$) and $\hat c_k^\dagger$ ($\hat c_k$)
create (annihilate) the electron at the molecular level $m$ or contact state $k$, respectively.
$\hat D^\dagger=\hat d_2^\dagger\hat d_1$ is the donor excitation operator.
$\hat a_\alpha^\dagger$ ($\hat a_\alpha$) creates (annihilates) excitation quanta
in the thermal bath.
We previously discussed a similar model~\cite{PeskinMGJCP12,WhitePeskinMGPRB13} 
in a  study of the quantum coherence effects on electron transport in DBA junctions. Einax et al.  \cite{EinaxDierlNitzanJPCC11,EinaxNitzanJPCC14}
utilized the model in consideration of a molecular heat engine within a hopping transport regime.
Here, we elucidate the effects of intra-molecular quantum coherence
(Rabi oscillation between the LUMOs, levels $2$ and $3$; see Figure~\ref{fig1})
on the thermodynamic performance of the engine.

The response properties of the junction can be conveniently expressed
in terms of NEGFs.
In particular, we use {single- } and two-particle electron Green functions
defined on the Keldysh contour as
 (here and below $e=\hbar=k_B=1$)
\begin{equation} 
\label{defG}
G_{m_1m_2}(\tau_1,\tau_2) = 
     -i\langle \mathcal{T}_c\,\hat d_{m_1}(\tau_1)\,\hat d_{m_2}^\dagger(\tau_2)\rangle;
\qquad
\mathcal{G}(\tau_1,\tau_2) = 
     -i\langle \mathcal{T}_c\, \hat D(\tau_1)\,\hat D^\dagger(\tau_2) \rangle.
\end{equation}
Here, $m_1$ and $m_2$ indicate the molecular levels, $\mathcal{T}_c$ is the contour ordering operator 
and $\tau_1$ and $\tau_2$ are the contour variables.
The single-particle Green function is obtained by solving the Dyson equation
\begin{equation}
 G_{m_1m_2}(\tau_1,\tau_2) = G^{(0)}_{m_1m_2}(\tau_1,\tau_2)
 + \sum_{m_3,m_4}\int_c d\tau_3\, \int_c d\tau_4\, 
 G^{(0)}_{m_1m_3}(\tau_1,\tau_3) \, \Sigma_{m_3m_4}(\tau_3,\tau_4)\, G_{m_4m_2}(\tau_4,\tau_2),
\end{equation} 
where $G^{(0)}_{m_1m_2}(\tau_1,\tau_2)$
is the single-particle Green function in the absence of coupling to baths and 
\begin{equation}
\label{SE}
\Sigma_{m_1m_2}(\tau_1,\tau_2)=\sum_{B=L,R,S} \Sigma^B_{m_1m_2}(\tau_1,\tau_2)
\end{equation}
is the total electron self-energy due to coupling to the baths.
In our consideration, we use second-order (in system-baths couplings) diagrammatic
expansion. The procedure is self-consistent, because single-electron self-energy
due to coupling to the radiation field $\Sigma^S_{m_1m_2}(\tau_1,\tau_2)$ depends on 
the single-electron
Green function $G_{m_1m_2}(\tau_1,\tau_2)$, while the latter is defined by the self-energy.
Below, we utilize level populations,
\begin{equation}
 n_m=-i G^{<}_{mm}(t,t)\quad (m=1,2,3),
\end{equation}
at subsequent iteration steps to judge the convergence of the procedure.
Here, $G^{<}$ is a lesser projection of the single-particle Green function.
Explicit expressions of the self-energies are given in Appendix~\ref{app_SE}.

To evaluate the two-particle Green function $\mathcal{G}$, we employ an approximation
\begin{equation}
 \mathcal{G}(\tau_1,\tau_2)\approx -i G_{11}(\tau_2,\tau_1)G_{22}(\tau_1,\tau_2)
\end{equation}
which disregards multi-photon processes in the evaluation of the heat flux (see below).
This approximation was employed in earlier studies~\cite{GalperinNitzanJCP06}, and for the parameters
of the simulations (strength of coupling to the radiation field), the approximation is reasonable.

Below, we calculate the particle flux at the interface with the right contact, $I_R$,
and the energy (heat) flux at the interface with the solar bath $S$, $J_S$.
In terms of these fluxes, the average thermodynamic efficiency of the molecular heat engine
is defined as
\begin{equation}
\label{etaTD}
 \eta = \frac{(\mu_R-\mu_L)I_R}{J_S} \equiv \frac{P}{J_S},
 \end{equation}
where $P\equiv(\mu_R-\mu_L)I_R$ is the power of the engine.
We are interested in the efficiency at the maximum power $\eta_{max}$.
Within NEGF, fluxes $I_R$ and $J_S$ are defined as the rates of the change 
in the electronic population in $R$ and the energy in $S$, respectively:
\begin{equation}
 I_R = -\frac{d}{dt}\sum_{r\in R} \langle \hat c_r^\dagger(t)\hat c_r(t)\rangle;
 \qquad
 J_S = -\frac{d}{dt}\sum_{\alpha\in S} \omega_\alpha \langle \hat a_\alpha^\dagger(t)\hat a_\alpha(t).\rangle
\end{equation}

They can be expressed exactly in terms of the Green functions and self-energies.
At steady state within the NEGF, fluxes $I_R$ and $J_S$ are
(see Appendix~\ref{app_FLX} for the derivation)
\begin{align}
 \label{IR}
 I_R =& \mbox{Tr}\int_{-\infty}^{+\infty}\frac{dE}{2\pi}
 \bigg(\Sigma^{R\, <}(E)\, G^{>}(E) - \Sigma^{R\, >}(E)\, G^{<}(E)\bigg),
 \\
 \label{JS}
 J_S =& -\int_0^{\infty}\frac{d\omega}{2\pi}\,\omega\,
 \bigg( \Pi^{<}(\omega)\, \mathcal{G}^{>}(\omega) - \Pi^{>}(\omega)\, \mathcal{G}^{<}(\omega) \bigg).
\end{align}

Here, {$\mbox{Tr}[\ldots]$ }is the trace over the molecular levels,
$\Sigma^{R\, <}(E)$ and $\Sigma^{R\, >}(E)$ are defined in Equations~(\ref{SKmp}) and (\ref{SKpm}),
$\Pi^{<}(\omega)$ and $\Pi^{>}(\omega)$ are given in Equations~(\ref{Pimp}) and (\ref{Pipm}).

%%%%%%%%%%%%%%%%%%%%%%%%/both%%%%%%%%%%%%%%%%%%
\section{Results}\label{numres}
We present the results of the simulations for the DBA heat engine model (see Figure~\ref{fig1}).
Unless stated otherwise, the parameters of the simulation are
ambient temperature $T=300$~K (this is the temperature of the contacts $L$ and $R$), 
temperature of the Sun $T_S=6000$~K (this is the temperature of the thermal bath $S$), 
molecular levels {$\varepsilon_{1}=-0.5$~eV and $\varepsilon_{2}=\varepsilon_3=0.8$~eV}, 
donor-acceptor electron hopping $t=0.1$~eV,  
contacts electron escape rates $\Gamma_L=\Gamma_R=0.01$~eV
and the energy dissipation rate to the thermal bath $\gamma=0.01$~eV.
Fermi energy is taken as the origin, $E_F=0$, and the junction is biased in contact $R$:
$\mu_L=E_F$ and $\mu_R=E_F+|e|V_{sd}$ (e is the electron charge). 
Here, $V_{sd}$ is the bias across the junction. Gate potential $V_g$ is applied to the acceptor; the position of the acceptor LUMO is $\varepsilon_3+|e|V_g$. Simulations are performed on the energy grid
spanning the region from $-$3 to 3~eV with step $5\times 10^{-4}$~eV.
Convergence is checked by comparing the level populations in the subsequent iterations
of the self-consistent solution of the Dyson equation. The convergence tolerance is $10^{-5}$.

\begin{figure}[htbp]
\centering\includegraphics[width=0.38\linewidth]{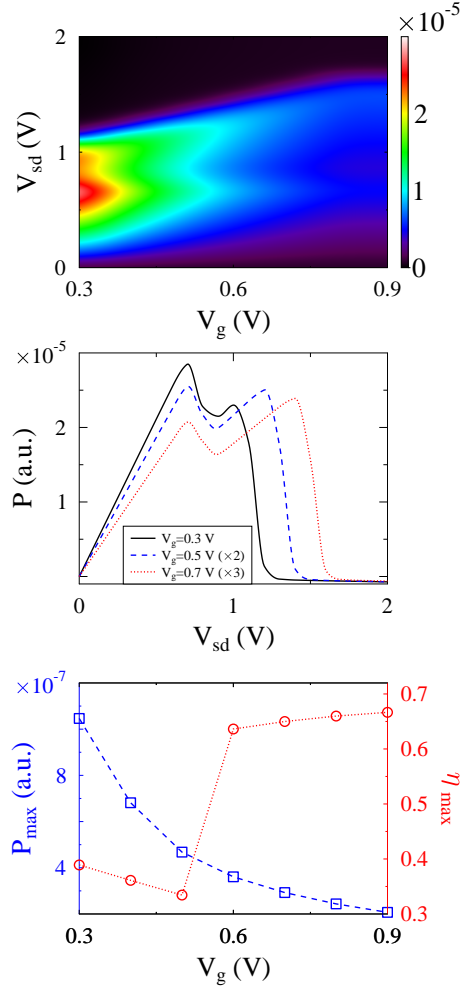}
\caption{\label{fig2}
Effects of coherence in the DBA heat engine (Figure~\ref{fig1}). 
Shown are (a) a map of the engine power $P$ vs. detuning of LUMOs, $V_g$,
and bias across the junction, $V_{sd}$; (b)  power $P$ vs. bias $V_{sd}$ for the three gate potentials
(see inset; the results for $V_g=0.5$ and $0.7$~V are scaled by a factor of $2$ and $3$, respectively); 
(c) the dependence of the maximum power $P_{max}$ 
(dashed line, squares) and the efficiency at maximum power $\eta_{max}$
(dotted line, circles) on gate potential $V_g$. 
}
\end{figure}

Figure~\ref{fig2} shows that quantum coherence between the LUMOs of the donor and acceptor
results in the two-peak structure of the engine power dependence on the bias at the fixed gate potential
(see the top panel). The effect can be easily understood if one transforms in the eigenbasis of 
the molecular system. In this basis, ground state ($\varepsilon_{1}$) is coupled by Sun radiation 
to two excited eigenstates. Each of the excited states yields a separate channel
for electron transfer into contact $R$. 
The strength of the excited states' coupling to the Sun radiation and the contact $R$ depends 
on gate potential $V_g$. Thus, the two-peak structure indicates the presence of two scattering 
eigenchannels with the dominant channel defined by detuning of the molecular system LUMOs
(see Figure~\ref{fig2}b). Note that this channel control is to some extent similar to
consideration of \cite{ScullyPRL10} with the role of the resonant driving field there
played by the gate potential in our consideration. Note also that channel control by 
level detuning does not affect the overall thermodynamic formulation, because
(for static levels) no external thermodynamic forces are acting on the system.

While maximum power (see Figure~\ref{fig2}c) is insensitive to the change
in the dominant scattering channel, the thermodynamic efficiency at maximum power
shows a non-monotonic behavior due to the sudden change in the bias at which the device
performs at maximum power when the dominant channel is switched.
We stress that this behavior is possible only due to the presence of quantum
coherence in the molecular system and that when coherence is destroyed the efficiency
at maximum power attains its monotonic dependence on level detuning
in agreement with the classical study of the system~\cite{EinaxDierlNitzanJPCC11}.

\begin{figure}[htbp]
\centering\includegraphics[width=0.38\linewidth]{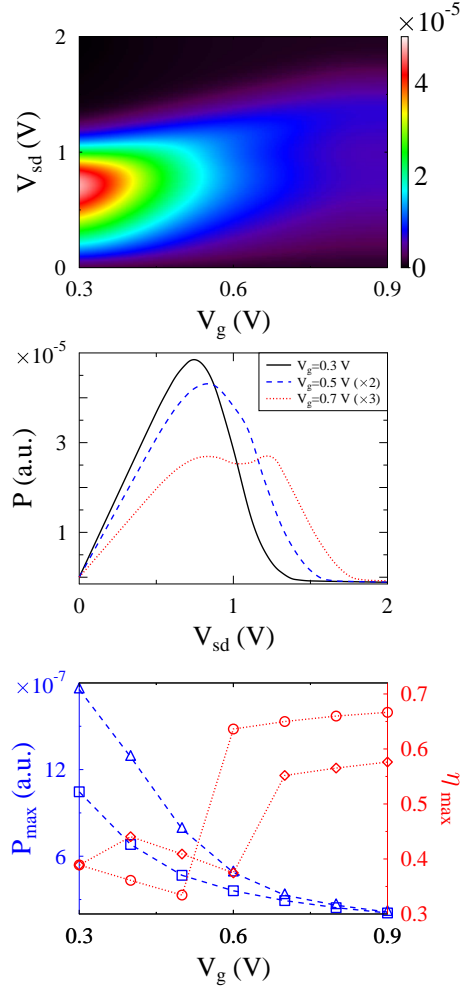}
\caption{\label{fig3}
Coherence destruction on efficiency of the DBA heat engine by a B{\" u}ttiker probe.
Shown are (a)  a map of the engine power $P$ vs. detuning of LUMOs, $V_g$,
and bias across the junction, $V_{sd}$; (b)  power $P$ vs. bias $V_{sd}$ for the three gate potentials
(see inset; the results for $V_g=0.5$ and $0.7$~V are scaled by a factor of $2$ and $3$, respectively);  
(c) the dependence of the maximum power $P_{max}$ 
(dashed line, triangles) and efficiency at maximum power $\eta_{max}$
(dotted line, diamonds) on gate potential $V_g$. 
The results presented in Fig.~\ref{fig2}c (coherent transport, $\delta=0$) --
squares (dashed line) and circles (dotted line) --
are given for comparison.
}
\end{figure}

To destroy system coherence, we employ B{\" u}ttiker probes coupled to LUMOs of the donor
and the acceptor. 
 {
B{\" u}ttiker probes are widely used in quantum transport literature.
They are modeled as an additional bath (represented by additional self-energy), whose role is to induce
dephasing in the system. The probes should destroy the phase
at the same time not allowing for either electron or energy exchange between the system and the probe.
}
Following~\cite{DattaPRB07}, we introduce the probes by considering
additional local self-energies
\begin{equation}
\label{deltadef}
 \Sigma_{mm}(\tau_1,\tau_2) = \delta\cdot G_{mm}(\tau_1,\tau_2)\quad (m=2,3).
\end{equation}
{Here}, $\delta$ is the dephasing parameter.
This self-energy can be derived considering the electron-phonon interaction in the second order of 
the diagrammatic expansion~\cite{HaugJauho_2008} in the limit of low phonon frequency~\cite{WhitePeskinMGPRB13}.
Such physical insight shows that neither electron nor energy flux
can be induced between the system and the probe. Thus, the probe can destroy only quantum coherence 
in the molecular system.   
%%%%%%%%%%%%%%%%%%%%%%%%%%%%%%%%%%%%%%%%%%%%%
%%%%%%%%%%%%%%%%%%%%%%%%%%%%%%%%%%%%%%%%%%%%%
Figure~\ref{fig3} demonstrates the effect of coherence destruction with B{\" u}ttiker probes.
The dephasing parameter employed in the simulation is $\delta = 0.02$~eV $^{2}$.
Introducing the probe has two effects on the device's performance.
First, by destroying coherence the two transport channel situation 
in the purely coherent case changes to a single transport channel situation in the purely classical
(hopping) case. The dephasing parameter employed in the simulation partially destroys
system coherence; therefore, the second channel is less prominent 
(compare panels (a) and (b) in Figures~\ref{fig2} and \ref{fig3}),
although the impact of coherence on efficiency is still pronounced
(compare the dotted lines in Figure~\ref{fig3}c).
Note that destroying coherence increases the efficiency for $V_{g}<0.6$~V, while decreasing it
for $V_g\geq 0.6$~V. This indicates the prevalence (in the purely coherent case) 
of destructive and constructive interference in the two regions.
Second, destroying coherence leads to a slight increase in the maximum power
(compare the dashed lines in Figure~\ref{fig3}c).   
The effect is due to the disruption by the B{\" u}ttiker probe of the Rabi oscillations between 
the donor and acceptor LUMOs: The electron that previously spent time in the system
now escapes faster in the right  contact, which results in an increase in the current.

%%%%%%%%%%%%%%%%%%%%%%%%%%%%%%%%%%%%%%%%%%%%%
%%%%%%%%%%%%%%%%%%%%%%%%%%%%%%%%%%%%%%%%%%%%%

Figure~\ref{fig3} demonstrates the effect of coherence destruction with B{\" u}ttiker probes.
The dephasing parameter employed in the simulation is $\delta=0.02$~eV $^{2}$.
Introducing the probe has two effects on the device's performance.
First, by destroying coherence the two transport channel situation 
in the purely coherent case changes to a single transport channel situation in the purely classical
(hopping) case. The dephasing parameter employed in the simulation partially destroys
system coherence; therefore, the second channel is less prominent 
(compare (a) and (b) panels in Figures~\ref{fig2} and \ref{fig3}),
although the impact of coherence on efficiency is still pronounced
(compare the dotted lines in the bottom panel in Figure~\ref{fig3}).
Note that destroying coherence increases the efficiency for $V_{g}<0.6$~V, while decreasing it
for $V_g\geq 0.6$~V. This indicates the prevalence (in the purely coherent case) 
of destructive and constructive interference in the two regions.
Second, destroying coherence leads to a slight increase in the maximum power
(compare the dashed lines in the bottom panel in Figure~\ref{fig3}).   
The effect is due to the disruption by the B{\" u}ttiker probe of the Rabi oscillations between 
the donor and acceptor LUMOs: The electron that previously spent time in the system
now escapes faster in the right  contact, which results in an increase in the current.

%%%%%%%%%%%%%%%%%%%%%%%%%%%%%%%%%%%%%%%%%%%%%
\begin{figure}[htbp]
\centering\includegraphics[width=0.38\linewidth]{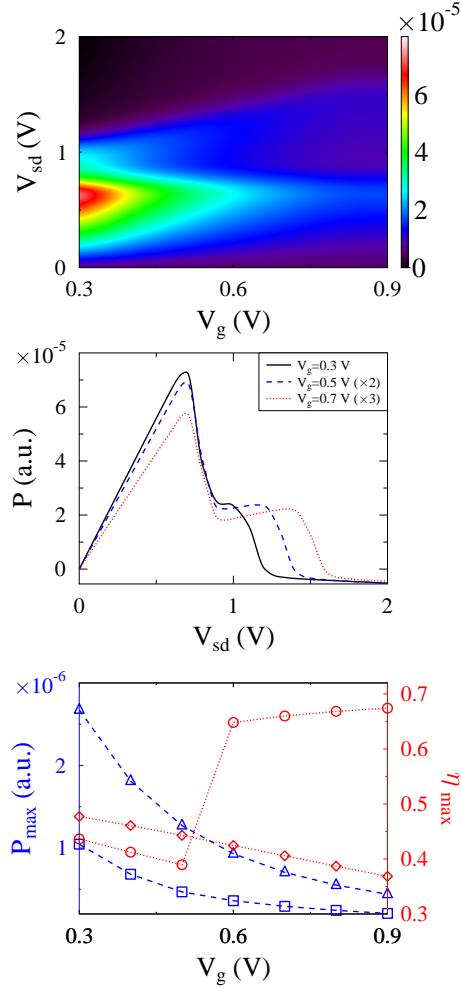}
\caption{\label{fig4}
Coherence destruction on efficiency of the DBA heat engine by dissipation due to coupling to contacts.
Shown are (a) a map of the engine power $P$ vs. detuning of the LUMOs, $V_g$,
and bias across the junction, $V_{sd}$; (b) power $P$ vs. bias $V_{sd}$ for the three gate potentials
(see inset; the results for $V_g=0.5$ and $0.7$~V are scaled by a factor of $2$ and $3$, respectively);
(c) the dependence of the maximum power $P_{max}$ 
(dashed line, triangles) and efficiency at maximum power $\eta_{max}$
(dotted line, diamonds) on gate potential $V_g$. 
The results presented in Figure~\ref{fig2}c{---}squares (dashed line) and circles (dotted line){---}are given for comparison.
}
\end{figure}
%%%%%%%%%%%%%%%%%%%%%%%%%%%%%%%%%%%%%%%%%%%%%

Another way to destroy system coherence is by increasing dissipation due to coupling to contacts
(increase escape rate parameters $\Gamma_L$ and $\Gamma_R$). 
We use $\Gamma_L=\Gamma_R=0.05$~eV in the simulations.
Similar to the B{\" u}tiiker probe case, here the destruction of coherence results in
a transition from two-channel to single-channel transport (and from quantum to classical behavior).
The difference between the two cases is that while the former is pure decoherence,
which results in a transition from coherent transport to hopping, the latter effectively eliminates
the acceptor LUMO by incorporating it into the right contact. 
Thus, the extreme of the B{\" u}tiiker probe is the classical model of \cite{EinaxDierlNitzanJPCC11}
while strong acceptor-contact coupling results in the classical consideration of \cite{EspositoPRB09}.
Figure~\ref{fig4} shows that for the parameters of the simulation coherence 
is completely destroyed, which results in a monotonic behavior of efficiency and 
an essential increase in the maximum power.

%%%%%%%%%%%%%%%%%%%%%%%%%%%%%%%%%%%%%%%%%%

%%%%%%%%%%%%%%%%%%%%%%%%%%%%%%%%%%%%%%%%%%
\section{Conclusions}\label{conclude}
We study thermoelectric properties of a bi-chromophoric DBA junction model
driven by solar radiation of the donor complex.
In particular, utilizing NEGF formalism, we calculate the efficiency
at the maximum power and elucidate the role of intra-molecular quantum coherence in the efficiency. 
Coherence in the junction is controlled by effective donor-acceptor coupling and by detuning
of molecular levels at the two nodes of the molecular complex.  
We show that quantum coherence results in non-monotonic behavior of efficiency at maximum 
power vs. level detuning. The observed sudden jump indicates a switch between dominant
scattering eigenchannel in our two-site model (see Figure~\ref{fig1}).
Although the non-monotonic behavior of the current-voltage
characteristic in junctions is well-known, thus far nobody 
has discussed this effect with respect to efficiency.

We stress that this behavior is possible only due to the presence of quantum
coherence in the molecular system and that when coherence is destroyed, efficiency
at maximum power attains its monotonic dependence on level detuning
in agreement with previously performed classical studies of the system.
To show the quantum-classical transition, we destroy quantum coherence in the system
by either employing a B{\" u}ttiker probe or increasing molecule-contacts coupling.
In both cases, destruction of coherence implies a transition from two transport channels to one
transport channel and from quantum to classical consideration.
We indicate that employing a B{\" u}ttiker probe results in pure dephasing with 
transition from coherent transport to hopping, thus reducing the quantum model
to the classical consideration of~\cite{EinaxDierlNitzanJPCC11}.
At the same time, increasing the strength of molecule-contacts coupling leads to dissipation
and effectively eliminates donor LUMO by incorporating it into the right contact.
This reduces our model to the classical consideration of~\cite{EspositoPRB09}.
Comparing the classical and quantum results, we find that quantum coherence 
may be advantageous or disadvantageous for the performance of a molecular
thermoelectric device (interference effects may lead to an increase or decrease in
the efficiency).

%%%%%%%%%%%%%%%%%%%%%%%%%%%%%%%%%%%%%%%%%%
%%%%%%%%%%%%%%%%%%%%%%%%%%%%%%%%%%%%%%%%%%
\begin{acknowledgments}
We thank Paul Brumer and Christian Van den Broeck for helpful comments.
This material is based upon work supported by the National Science Foundation under CHE-1565939.
\end{acknowledgments}

\appendix
\section{Expressions for Self-Energies}\label{app_SE}
Expressions for self-energies Equation (\ref{SE}) can be derived following the standard diagrammatic perturbation
theory formulated on the Keldysh contour~\cite{FetterWalecka_1971,HaugJauho_2008,StefanucciVanLeeuwen_2013}.
The resulting expression for the electron self-energies due to coupling to contacts $L$ and $R$
is exact
\begin{equation}
 \label{defSK}
 \Sigma^{K}_{m_1m_2}(\tau_1,\tau_2) = \sum_{k\in K} V_{m_1k}\, g_k(\tau_1,\tau_2)\, V_{km_2}
 \quad (K=L,R)
\end{equation}
where
\begin{equation}
 g_k(\tau_1,\tau_2) \equiv -i\langle \mathcal{T}_c\,\hat c_l(\tau_1)\,\hat c_k^\dagger(\tau_2)\rangle
\end{equation}
is the Green function of the free electron in state $k$.
For {Equation} (\ref{Hexp}), $V_{mk}=V_\ell$ for $m=1$ and $V_{mk}=V_r$ for $m=3$.
At steady state and within the wide-band approximation, the projections of (\ref{defSK})
after performing the Fourier transform are
\begin{align}
 \label{SKmm}
 \Sigma_{m_1m_2}^{K\, --}(E) =&\, \Sigma_{m_1m_2}^{K\, ++}(E) = i\,\Gamma^K_{m_1m_2}\bigg(f_K(E)-\frac{1}{2}\bigg)
 \\
 \label{SKmp}
 \Sigma_{m_1m_2}^{K\, -+}(E) =&\, i\, \Gamma^K_{m_1m_2} \, f_K(E)
 \\
 \label{SKpm}
 \Sigma_{m_1m_2}^{K\, +-}(E) =& -i\, \Gamma^K_{m_1m_2} \bigg(1-f_K(E)\bigg),
\end{align}
where $K=L,R$, superscript $-$ ($+$) indicates the forward (backward) branches of the Keldysh contour,
$f_K(E)$ is the Fermi-Dirac distribution, and
\begin{equation}
 \Gamma^K_{m_1m_2}(E) \equiv 2\pi\sum_{k\in K} V_{m_1k}V_{km_2}\delta(E-\varepsilon_k)
\end{equation} 
is the dissipation matrix due to coupling to contacts. 
In the wide-band approximation, the matrix is energy-independent.
For Equation (\ref{Hexp}),
$\Gamma ^L_{m_1m_2}= \delta_{m_1,m_2,1}\, \Gamma_L$ and
$\Gamma ^R_{m_1m_2}= \delta_{m_1,m_2,3}\, \Gamma_R$,
where $\Gamma_L$ and $\Gamma_R$ are the electron escape rates into contacts $L$ and $R$,
respectively. Note $\Sigma^{K\, <(>)}\equiv\Sigma^{K\, -+(+-)}$.

The expression for the self-energy due to coupling to the thermal bath is derived within
the second order of the diagrammatic perturbation theory
\begin{equation}
 \label{defSS}
 \Sigma^S_{m_1m_2}(\tau_1,\tau_2) = 
 \begin{cases}
 i\, \Pi(\tau_2,\tau_1)\, G_{22}(\tau_1,\tau_2) & \mbox{for}\  m_1=m_2=1
 \\
 i\, \Pi(\tau_1,\tau_2)\, G_{11}(\tau_1,\tau_2) & \mbox{for}\  m_1=m_2=2
 \\
 0 & \mbox{otherwise},
 \end{cases}
\end{equation}
where
\begin{equation}
 \label{defPi}
 \Pi(\tau_1,\tau_2) \equiv \sum_\alpha \lvert U_\alpha\rvert^2 F_\alpha(\tau_1,\tau_2)
\end{equation}
is the self-energy due to the coupling of the molecular excitations to the thermal bath.
The projections after the Fourier transform is performed are
\begin{equation}
 \Sigma^{S\, s_1s_2}_{m_1m_2}(E) = i\int_0^{+\infty}\frac{d\omega}{2\pi}
 \begin{cases}
  \Pi^{s_2s_1}(\omega)\, G^{s_1s_2}_{22}(E+\omega) & \mbox{for}\  m_1=m_2=1
  \\
   \Pi^{s_1s_2}(\omega)\, G^{s_1s_2}_{22}(E-\omega) & \mbox{for}\  m_1=m_2=2
    \\
 0 & \mbox{otherwise}.
 \end{cases}
\end{equation}
{Here}, $s_{1,2}=\{-,+\}$ and
\begin{align}
 \label{Pimm}
 \Pi^{--}(E) =&\, \Pi^{++}(E) = -i\,\gamma\bigg(N(E)+\frac{1}{2}\bigg)
 \\
 \label{Pimp}
 \Pi^{-+}(E) =& -i\, \gamma \, N(E)
 \\
 \label{Pipm}
 \Pi^{+-}(E) =& -i\, \gamma \bigg(1+N(E)\bigg),
\end{align}
$N(E)$ is the Bose-Einstein distribution, and
\begin{equation}
 \label{defgamma}
 \gamma(\omega)\equiv 2\pi\sum_{\alpha}\lvert U_\alpha\rvert^2\delta(\omega-\omega_\alpha)
\end{equation}
is the energy dissipation rate.

%%%%%%%%%%%%%%%%%%%%%%%%%%%%%%%%%%%%%%%%%%

\section{Derivation of Fluxes}\label{app_FLX}
The expression for electron flux, Equation (\ref{IR}), is the celebrated Jauho, Wingreen and Meir 
formula \cite{JauhoWingreenMeirPRB94} written for the steady-state situation.

Here, we will focus on the derivation of the photon energy flux Equation~(\ref{JS}).
We start from the definition of the flux as the rate of the change of energy in the thermal bath 
(the radiation field)
\begin{equation}
 J_S(t) \equiv -\frac{d}{dt} \sum_\alpha \omega_\alpha \langle \hat a_\alpha^\dagger(t)\hat a_\alpha(t)\rangle .
\end{equation}
Here, {$\langle\ldots\rangle$} is the statistical and quantum mechanical averaging with 
the density operator
of the whole world (system plus baths), and the operators are represented in the Heisenberg picture.
Using the Heisenberg equation-of-motion (EOM) for the operators $\hat a_\alpha^\dagger(t)$
and $\hat a_\alpha(t)$ within Equation (\ref{Hexp}), one arrives at the expression for the flux 
\begin{equation}
 \label{step1}
 J_S(t) = 2\mbox{Re}\sum_\alpha\omega_\alpha U_\alpha G^{<}_{\alpha D}(t,t)
\end{equation}
in terms of the lesser projection of a mixed (photon-molecular excitation) Green function
\begin{equation}
 G_{\alpha D}(\tau_1,\tau_2) \equiv -i\langle \mathcal{T}_c\,\hat a_\alpha(\tau_1)\,\hat D^\dagger(\tau_2)\rangle.
\end{equation}
Writing the Dyson equation for the Green function in the integral form and taking a lesser projection
leads to
\begin{equation}
 \label{GaD}
 G_{\alpha D}^{<}(t,t) = \int_{-\infty}^{+\infty} dt_1\, \bigg( F_\alpha^{<}(t,t_1)\, U_\alpha^{*}\,\mathcal{G}^{a}(t_1,t)
 + F_{\alpha}^{r}(t,t_1)\, U_\alpha^{*}\,\mathcal{G}^{<}(t_1,t)\bigg),
\end{equation}
where $\mathcal{G}$ is the two-particle Green function defined in Equation~(\ref{defG}) and
\begin{equation}
 F_\alpha(\tau_1,\tau_2) \equiv -i\langle \mathcal{T}_c\, \hat a_\alpha(\tau_1)\,\hat a_\alpha^\dagger(\tau_2)\rangle
\end{equation}
is the Green function of the free photon of the radiation field.
Substituting Equation~(\ref{GaD}) into Equation~(\ref{step1}) 
and using
\begin{align}
 F_\alpha^r(t,t_1) \equiv& \theta(t-t_1)\bigg(F_\alpha^{>}(t,t_1)-F_\alpha^{<}(t,t_1)\bigg)
 \\
 \mathcal{G}^a(t_1,t) \equiv& \theta(t-t_1)\bigg(\mathcal{G}^{<}(t_1,t)-\mathcal{G}^{>}(t_1,t)\bigg) 
\end{align}
(here, $\theta(x)$ is the Heaviside step function) leads to 
\begin{equation}
 \label{step2}
 J_S(t) = -2\,\mbox{Re} \sum_\alpha\lvert U_\alpha\rvert^2 \omega_\alpha 
 \int_{-\infty}^{t} dt_1\bigg( \partial_t F_\alpha^{<}(t,t_1)\,\mathcal{G}^{>}(t_1,t)
 + \partial_t F_\alpha^{>}(t,t_1)\,\mathcal{G}^{<}(t_1,t) \bigg).
\end{equation}
Finally, taking into account that for the free photon in mode $\alpha$
$F_\alpha^{<(>)}(t,t_1)\sim e^{-i\omega_\alpha(t-t_1)}$, 
and thus, $\omega_\alpha F_\alpha^{<(>)}(t,t_1)\equiv i\partial_t F_\alpha^{<(>)}(t,t_1)$,
we get
\begin{equation}
 \label{step3}
 J_S(t) = -2i\,\mbox{Re} \int_{-\infty}^{t} dt_1\bigg( \partial_t\Pi^{<}(t,t_1)\,\mathcal{G}^{>}(t_1,t)
 - \partial_t\Pi^{>}(t,t_1)\,\mathcal{G}^{<}(t_1,t) \bigg),
\end{equation}
where $\Pi(\tau_1,\tau_2)$ is defined in Equation~(\ref{defPi}). 
At steady-state, the correlation functions depend only on the time difference; 
therefore, performing the Fourier transform in Equation~(\ref{step3}) leads  to Equation~(\ref{JS}).

%%%%%%%%%%%%%%%%%%%%%%%%%%%%%%%%%%%%%%%%%%

\begin{thebibliography}{-------}
\providecommand{\natexlab}[1]{#1}

\bibitem[Aviram and Ratner(1974)]{AviramRatnerCPL74}
Aviram, A.; Ratner, M.A.
\newblock Molecular Rectifiers.
\newblock {\em Chem. Phys. Lett.} {\bf 1974}, {\em 29},~277--283.

\bibitem[Ratner(2013)]{RatnerNatNano13}
Ratner, M.
\newblock A Brief History of Molecular Electronics.
\newblock {\em Nat. Nanotech.} {\bf 2013}, {\em 8},~378--381.

\bibitem[van~der Molen \em{et~al.}(2013)van~der Molen, Naaman, Scheer, Neaton,
  Nitzan, Natelson, Tao, van~der Zant, Mayor, Ruben, Reed, and
  Calame]{visionsNatNano13} van~der Molen, S.J.; Naaman, R.; Scheer, E.; Neaton, J.B.; Nitzan, A.;
Natelson, D.; Tao, N.J.; van~der Zant, H.S.J.; Mayor, M.; Ruben, M.; Reed,
M.; Calame, M.
\newblock Visions for a Molecular Future.
\newblock {\em Nat. Nanotech.} {\bf 2013}, {\em 8},~385--389.%%please check.

\bibitem[Hensley \em{et~al.}(2012)Hensley, Yang, and Centurion]{CenturionPRL12}
Hensley, C.J.; Yang, J.; Centurion, M.
\newblock Imaging of Isolated Molecules with Ultrafast Electron Pulses.
\newblock {\em Phys. Rev. Lett.} {\bf 2012}, {\em 109},~133202.

\bibitem[Ham and Ho(2013)]{HoJCP13}
Ham, U.; Ho, W.
\newblock Imaging Single Electron Spin in a Molecule Trapped within a
  Nanocavity of Tunable Dimension.
\newblock {\em J. Chem. Phys.} {\bf 2013}, {\em 138},~074703.

\bibitem[Schaffert \em{et~al.}(2013)Schaffert, Cottin, Sonntag, Karacuban,
  Bobisch, Lorente, Gauyacq, and M{\" o}ller]{SchaffertNatMater13}
Schaffert, J.; Cottin, M.C.; Sonntag, A.; Karacuban, H.; Bobisch, C.A.;
  Lorente, N.; Gauyacq, J.P.; M{\" o}ller, R.
\newblock Imaging the Dynamics of Individually Adsorbed Molecules.
\newblock {\em Nat. Mater.} {\bf 2013}, {\em 12},~223--227.

\bibitem[Zhang \em{et~al.}(2013)Zhang, Zhang, Dong, Jiang, Zhang, Chen, Zhang,
  Liao, Aizpurua, Luo, Yang, and Hou]{DongHouNature13}
Zhang, R.; Zhang, Y.; Dong, Z.C.; Jiang, S.; Zhang, C.; Chen, L.G.; Zhang, L.;
  Liao, Y.; Aizpurua, J.; Luo, Y.; Yang, J.L.; Hou, J.G.
\newblock Chemical Mapping of a Single Molecule by Plasmon-Enhanced Raman
  Scattering.
\newblock {\em Nature} {\bf 2013}, {\em 498},~82--86.

\bibitem[Chiang \em{et~al.}(2014)Chiang, Xu, Han, and Ho]{HoScience14}
Chiang, C.l.; Xu, C.; Han, Z.; Ho, W.
\newblock Real-Space Imaging of Molecular Structure and Chemical Bonding by
  Single-Molecule Inelastic Tunneling Probe.
\newblock {\em Science} {\bf 2014}, {\em 344},~885--888.

\bibitem[Schwarz \em{et~al.}(2015)Schwarz, Kastlunger, Lissel, Egler-Lucas,
  Semenov, Venkatesan, Berke, Stadler, and L{\" o}rtscher]{LortscherNatNano16}
Schwarz, F.; Kastlunger, G.; Lissel, F.; Egler-Lucas, C.; Semenov, S.N.;
  Venkatesan, K.; Berke, H.; Stadler, R.; L{\" o}rtscher, E.
\newblock Field-Induced Conductance Switching by Charge-State Alternation in
  Organometallic Single-Molecule Junctions.
  \newblock {\em Nat. Nanotech.} {\bf 2016}, {\em 11},~170--176.
\bibitem[Mayor \em{et~al.}(2003)Mayor, Weber, Reichert, Elbing, von
  H{\"{a}}nisch, Beckmann, and Fischer]{MayorWeberAngChim03}
Mayor, M.; Weber, H.B.; Reichert, J.; Elbing, M.;{von H{\"{a}}nisch, C.};
  Beckmann, D.; Fischer, M.
\newblock Electric Current through a Molecular Rod-Relevance of the Position
  of the Anchor Groups.
\newblock {\em Angew. Chem. Int. Ed.} {\bf 2003}, {\em 42},~5834--5838.

\bibitem[Vazquez \em{et~al.}(2012)Vazquez, Skouta, Schneebeli, Kamenetska,
  Breslow, Venkataraman, and Hybertsen]{VenkataramanHybertsenNatNano12}
Vazquez, H.; Skouta, R.; Schneebeli, S.; Kamenetska, M.; Breslow, R.;
  Venkataraman, L.; Hybertsen, M.
\newblock Probing the Conductance Superposition Law in Single-Molecule Circuits
  with Parallel Paths.
\newblock {\em Nature Nanotech.} {\bf 2012}, {\em 7},~663--667.

\bibitem[Ballmann \em{et~al.}(2012)Ballmann, H\"artle, Coto, Elbing, Mayor,
  Bryce, Thoss, and Weber]{WeberPRL12}
Ballmann, S.; H\"artle, R.; Coto, P.B.; Elbing, M.; Mayor, M.; Bryce, M.R.;
  Thoss, M.; Weber, H.B.
\newblock Experimental Evidence for Quantum Interference and Vibrationally
  Induced Decoherence in Single-Molecule Junctions.
\newblock {\em Phys. Rev. Lett.} {\bf 2012}, {\em 109},~056801.

\bibitem[Frontiera \em{et~al.}(2012)Frontiera, Gruenke, and
  Duyne]{VanDuyneNL12}
Frontiera, R.R.; Gruenke, N.L.; Duyne, R.P.V.
\newblock Fano-Like Resonances Arising from Long-Lived Molecule-Plasmon
  Interactions in Colloidal Nanoantennas.
\newblock {\em Nano Lett.} {\bf 2012}, {\em 12},~5989--5994.

\bibitem[Lee \em{et~al.}(2014)Lee, Perdue, Perez, and
  Apkarian]{ApkarianACSNano14}
Lee, J.; Perdue, S.M.; Perez, A.R.; Apkarian, V.A.
\newblock Vibronic Motion with Joint Angstrom-Femtosecond Resolution Observed
  through Fano Progressions Recorded within One Molecule.
\newblock {\em ACS Nano} {\bf 2014}, {\em 8},~54--63.

\bibitem[Lee \em{et~al.}(2007)Lee, Cheng, and Fleming]{FlemingScience07}
Lee, H.; Cheng, Y.C.; Fleming, G.R.
\newblock Coherence Dynamics in Photosynthesis: Protein Protection of Excitonic
  Coherence.
\newblock {\em Science} {\bf 2007}, {\em 316},~1462--1465.

\bibitem[Engel \em{et~al.}(2007)Engel, Calhoun, Read, Ahn, Mancal, Cheng,
  Blankenship, and Fleming]{EngelFlemingNature07}
Engel, G.S.; Calhoun, T.R.; Read, E.L.; Ahn, T.K.; Mancal, T.; Cheng, Y.C.;
  Blankenship, R.E.; Fleming, G.R.
\newblock Evidence for Wavelike Energy Transfer through Quantum Coherence in
  Photosynthetic Systems.
\newblock {\em Nature} {\bf 2007}, {\em 446},~782--786.

\bibitem[Panitchayangkoon \em{et~al.}(2010)Panitchayangkoon, Hayes, Fransted,
  Caram, Harel, Wen, Blankenship, and Engel]{EngelPNAS10}
Panitchayangkoon, G.; Hayes, D.; Fransted, K.A.; Caram, J.R.; Harel, E.; Wen,
  J.; Blankenship, R.E.; Engel, G.S.
\newblock Long-Lived Quantum Coherence in Photosynthetic Complexes at
  Physiological Temperature.
\newblock {\em Proc. Natl. Acad. Sci.} {\bf 2010}, {\em 107},~12766--12770.

\bibitem[Ishizaki \em{et~al.}(2010)Ishizaki, Calhoun, Schlau-Cohen, and
  Fleming]{FlemingPCCP10}
Ishizaki, A.; Calhoun, T.R.; Schlau-Cohen, G.S.; Fleming, G.R.
\newblock Quantum Coherence and its Interplay with Protein Environments in
  Photosynthetic Electronic Energy Transfer.
\newblock {\em Phys. Chem. Chem. Phys.} {\bf 2010}, {\em 12},~7319--7337.

\bibitem[Ioffe(1957)]{Ioffe_1957}
Ioffe, A.F.
\newblock {\em Semiconductor Thermoelements and Thermoelectric Cooling};
  Infosearch Ltd.: London, UK, 1957.

\bibitem[Kim \em{et~al.}(2001)Kim, Shi, Majumdar, and
  McEuen]{MajumdarMcEuenPRL01}
Kim, P.; Shi, L.; Majumdar, A.; McEuen, P.L.
\newblock Thermal Transport Measurements of Individual Multiwalled Nanotubes.
\newblock {\em Phys. Rev. Lett.} {\bf 2001}, {\em 87},~215502.

\bibitem[Reddy \em{et~al.}(2007)Reddy, Jang, Segalman, and
  Majumdar]{MajumdarScience07}
Reddy, P.; Jang, S.Y.; Segalman, R.A.; Majumdar, A.
\newblock Thermoelectricity in Molecular Junctions.
\newblock {\em Science} {\bf 2007}, {\em 315},~1568--1571.

\bibitem[Malen \em{et~al.}(2010)Malen, Yee, Majumdar, and
  Segalman]{MajumdarCPL10}
Malen, J.A.; Yee, S.K.; Majumdar, A.; Segalman, R.A.
\newblock Fundamentals of Energy Transport, Energy Conversion, and Thermal
  Properties in Organic-Inorganic Heterojunctions.
\newblock {\em Chem. Phys. Lett.} {\bf 2010}, {\em 491},~109--122.

\bibitem[Widawsky \em{et~al.}(2012)Widawsky, Darancet, Neaton, and
  Venkataraman]{VenkataramanNL12}
Widawsky, J.R.; Darancet, P.; Neaton, J.B.; Venkataraman, L.
\newblock Simultaneous Determination of Conductance and Thermopower of Single
  Molecule Junctions.
\newblock {\em Nano Lett.} {\bf 2012}, {\em 12},~354--358.

\bibitem[Tsutsui \em{et~al.}(2013)Tsutsui, Morikawa, Arima, and
  Taniguchi]{TsutsuiSciRep13}
Tsutsui, M.; Morikawa, T.; Arima, A.; Taniguchi, M.
\newblock Thermoelectricity in Atom-Sized Junctions at Room Temperatures.
\newblock {\em Sci. Rep.} {\bf 2013}, {\em 3},~3326.

\bibitem[Brantut \em{et~al.}(2013)Brantut, Grenier, Meineke, Stadler, Krinner,
  Kollath, Esslinger, and Georges]{EsslingerScience13}
Brantut, J.P.; Grenier, C.; Meineke, J.; Stadler, D.; Krinner, S.; Kollath, C.;
  Esslinger, T.; Georges, A.
\newblock A Thermoelectric Heat Engine with Ultracold Atoms.
\newblock {\em Science} {\bf 2013}, {\em 342},~713--715.

\bibitem[Chabinyc(2014)]{ChabinycNatMater14}
Chabinyc, M.
\newblock Thermoelectric Polymers: Behind Organics' Thermopower.
\newblock {\em Nature Mater.} {\bf 2014}, {\em 13},~119--121.

\bibitem[Kim \em{et~al.}(2014)Kim, Jeong, Kim, Lee, and Reddy]{ReddyNatNano14}
Kim, Y.; Jeong, W.; Kim, K.; Lee, W.; Reddy, P.
\newblock Electrostatic Control of Thermoelectricity in Molecular Junctions.
\newblock {\em Nature Nanotech.} {\bf 2014}, {\em 9},~881--885.

\bibitem[Lee \em{et~al.}(2014)Lee, Cho, Lyeo, and Kim]{LeePRL14}
Lee, E.S.; Cho, S.; Lyeo, H.K.; Kim, Y.H.
\newblock Seebeck Effect at the Atomic Scale.
\newblock {\em Phys. Rev. Lett.} {\bf 2014}, {\em 112},~136601.

\bibitem[Paulsson and Datta(2003)]{PaulssonDattaPRB03}
Paulsson, M.; Datta, S.
\newblock Thermoelectric Effect in Molecular Electronics.
\newblock {\em Phys. Rev. B} {\bf 2003}, {\em 67},~241403.

\bibitem[Humphrey and Linke(2005)]{LinkePRL05}
Humphrey, T.E.; Linke, H.
\newblock Reversible Thermoelectric Nanomaterials.
\newblock {\em Phys. Rev. Lett.} {\bf 2005}, {\em 94},~096601.

\bibitem[Galperin \em{et~al.}(2008)Galperin, Nitzan, and
  Ratner]{MGNitzanRatnerMP08}
Galperin, M.; Nitzan, A.; Ratner, M.A.
\newblock Inelastic Effects in Molecular Junction Transport: Scattering and
  Self-Consistent Calculations for the Seebeck Coefficient.
\newblock {\em Mol. Phys.} {\bf 2008}, {\em 106},~397--404.

\bibitem[Liu \em{et~al.}(2009)Liu, Chen, and Chen]{LiuChenACSNano09}
Liu, Y.S.; Chen, Y.R.; Chen, Y.C.
\newblock Thermoelectric Efficiency in Nanojunctions: A Comparison between
  Atomic Junctions and Molecular Junctions.
\newblock {\em ACS Nano} {\bf 2009}, {\em 3},~3497--3504.

\bibitem[Liu and Chen(2009)]{LiuChenPRB09}
Liu, Y.S.; Chen, Y.C.
\newblock Seebeck Coefficient of Thermoelectric Molecular Junctions:
  First-Principles Calculations.
\newblock {\em Phys. Rev. B} {\bf 2009}, {\em 79},~193101.

\bibitem[Dubi and Di~Ventra(2009)]{DubiDiVentraNL09}
Dubi, Y.; Di~Ventra, M.
\newblock Thermoelectric Effects in Nanoscale Junctions.
\newblock {\em Nano Lett.} {\bf 2009}, {\em 9},~97--101.

\bibitem[Leijnse \em{et~al.}(2010)Leijnse, Wegewijs, and
  Flensberg]{WegewijsFlensbergPRB10}
Leijnse, M.; Wegewijs, M.R.; Flensberg, K.
\newblock Nonlinear Thermoelectric Properties of Molecular Junctions with
  Vibrational Coupling.
\newblock {\em Phys. Rev. B} {\bf 2010}, {\em 82},~045412.

\bibitem[Fransson and Galperin(2011)]{FranssonMGPCCP11}
Fransson, J.; Galperin, M.
\newblock Spin Seebeck Coefficient of a Molecular Spin Pump.
\newblock {\em Phys. Chem. Chem. Phys.} {\bf 2011}, {\em 13},~14350--14357.

\bibitem[Entin-Wohlman \em{et~al.}(2010)Entin-Wohlman, Imry, and
  Aharony]{EntinWohlmanImryAharonyPRB10}
Entin-Wohlman, O.; Imry, Y.; Aharony, A.
\newblock Three-Terminal Thermoelectric Transport through a Molecular Junction.
\newblock {\em Phys. Rev. B} {\bf 2010}, {\em 82},~115314.

\bibitem[Entin-Wohlman and Aharony(2012)]{EntinWohlmanAharonyPRB12}
Entin-Wohlman, O.; Aharony, A.
\newblock Three-Terminal Thermoelectric Transport under Broken Time-Reversal
  Symmetry.
\newblock {\em Phys. Rev. B} {\bf 2012}, {\em 85},~085401.

\bibitem[Nikoli{\' c} \em{et~al.}(2012)Nikoli{\' c}, Saha, Markussen, and
  Thygesen]{ThygesenJCompEl12}
Nikoli{\' c}, B.K.; Saha, K.K.; Markussen, T.; Thygesen, K.S.
\newblock First-Principles Quantum Transport Modeling of Thermoelectricity in
  Single-Molecule Nanojunctions with Graphene Nanoribbon Electrodes.
\newblock {\em J. Comput. Electron.} {\bf 2012}, {\em 11},~78--92.

\bibitem[S\'anchez and L\'opez(2013)]{SanchezLopezPRL13}
S\'anchez, D.; L\'opez, R.
\newblock Scattering Theory of Nonlinear Thermoelectric Transport.
\newblock {\em Phys. Rev. Lett.} {\bf 2013}, {\em 110},~026804.

\bibitem[Rutten \em{et~al.}(2009)Rutten, Esposito, and Cleuren]{EspositoPRB09}
Rutten, B.; Esposito, M.; Cleuren, B.
\newblock Reaching Optimal Efficiencies Using Nanosized Photoelectric Devices.
\newblock {\em Phys. Rev. B} {\bf 2009}, {\em 80},~235122.

\bibitem[Einax \em{et~al.}(2011)Einax, Dierl, and
  Nitzan]{EinaxDierlNitzanJPCC11}
Einax, M.; Dierl, M.; Nitzan, A.
\newblock Heterojunction Organic Photovoltaic Cells as Molecular Heat Engines:
  A Simple Model for the Performance Analysis.
\newblock {\em J. Phys. Chem. C} {\bf 2011}, {\em 115},~21396--21401.

\bibitem[Creatore \em{et~al.}(2013)Creatore, Parker, Emmott, and
  Chin]{CreatorePRL13}
Creatore, C.; Parker, M.A.; Emmott, S.; Chin, A.W.
\newblock Efficient Biologically Inspired Photocell Enhanced by Delocalized
  Quantum States.
\newblock {\em Phys. Rev. Lett.} {\bf 2013}, {\em 111},~253601.

\bibitem[Einax and Nitzan(2014)]{EinaxNitzanJPCC14}
Einax, M.; Nitzan, A.
\newblock Network Analysis of Photovoltaic Energy Conversion.
\newblock {\em J. Phys. Chem. C} {\bf 2014}, {\em 118},~27226--27234.

\bibitem[Einax and Nitzan(2015)]{EinaxNitzanARXIV15}
Einax, M.; Nitzan, A.
\newblock Maximum Efficiency of State-Space Models of Molecular Scale Engines.
\newblock arxiv:1506.00496,  2015.

\bibitem[Rahav \em{et~al.}(2012)Rahav, Harbola, and
  Mukamel]{HarbolaMukamelPRA12}
Rahav, S.; Harbola, U.; Mukamel, S.
\newblock Heat Fluctuations and Coherences in a Quantum Heat Engine.
\newblock {\em Phys. Rev. A} {\bf 2012}, {\em 86},~043843.

\bibitem[Dorfman \em{et~al.}(2013)Dorfman, Voronine, Mukamel, and
  Scully]{MukamelPNAS13}
Dorfman, K.E.; Voronine, D.V.; Mukamel, S.; Scully, M.O.
\newblock Photosynthetic reaction center as a quantum heat engine.
\newblock{\em Proc.  Natl. Acad.  Sci. U.S.A.} {\bf 2013},
  {\em 110},~2746--2751.

\bibitem[Goswami and Harbola(2013)]{HarbolaPRA13}
Goswami, H.P.; Harbola, U.
\newblock Thermodynamics of Quantum Heat Engines.
\newblock {\em Phys. Rev. A} {\bf 2013}, {\em 88},~013842.

\bibitem[Altintas \em{et~al.}(2015)Altintas, Hardal, and M\"ustecaplio{\v
  g}lu]{MustecapliogluPRA15}
Altintas, F.; Hardal, A.U.C.; M\"ustecaplio{\v g}lu, O.E.
\newblock Rabi Model as a Quantum Coherent Heat Engine: From Quantum Biology to
  Superconducting Circuits.
\newblock {\em Phys. Rev. A} {\bf 2015}, {\em 91},~023816.

\bibitem[Leggio \em{et~al.}(2015)Leggio, Bellomo, and Antezza]{LeggioPRA15}
Leggio, B.; Bellomo, B.; Antezza, M.
\newblock Quantum Thermal Machines with Single Nonequilibrium Environments.
\newblock {\em Phys. Rev. A} {\bf 2015}, {\em 91},~012117.

\bibitem[Leijnse and Wegewijs(2008)]{LeijnseWegewijsPRB08}
Leijnse, M.; Wegewijs, M.R.
\newblock Kinetic equations for transport through single-molecule transistors.
\newblock {\em Phys. Rev. B} {\bf 2008}, {\em 78},~235424.

\bibitem[Esposito and Galperin(2010)]{EspositoMGJPCC10}
Esposito, M.; Galperin, M.
\newblock Self-Consistent Quantum Master Equation Approach to Molecular
  Transport.
\newblock {\em J. Phys. Chem. C} {\bf 2010}, {\em 114},~20362--20369.

\bibitem[Esposito \em{et~al.}(2015)Esposito, Ochoa, and Galperin]{EspMGPRB15}
Esposito, M.; Ochoa, M.A.; Galperin, M.
\newblock Efficiency fluctuations in quantum thermoelectric devices.
\newblock {\em Phys. Rev. B} {\bf 2015}, {\em 91},~115417.

\bibitem[Gao and Galperin(2016)]{GaoMGJCP16_2}
Gao, Y.; Galperin, M.
\newblock Simulation of optical response functions in molecular junctions.
\newblock {\em J. Chem. Phys.} {\bf 2016}, {\em 144},~244106.

\bibitem[Kosloff and Levy(2014)]{KosloffLevyAnnRevPhysChem14}
Kosloff, R.; Levy, A.
\newblock Quantum Heat Engines and Refrigerators: Continuous Devices.
\newblock {\em Ann. Rev. Phys. Chem.} {\bf 2014}, {\em 65},~365--393.

\bibitem[Proesmans \em{et~al.}(2015)Proesmans, Cleuren, and den
  Broeck]{VanDenBroeckEPL15}
Proesmans, K.; Cleuren, B.; den Broeck, C.V.
\newblock Stochastic efficiency for effusion as a thermal engine.
\newblock {\em Europhys. Lett.} {\bf 2015}, {\em 109},~20004.

\bibitem[Kosloff and Rezek(2017)]{KosloffEntropy17}
Kosloff, R.; Rezek, Y.
\newblock The Quantum Harmonic Otto Cycle.
\newblock {\em Entropy} {\bf 2017}, {\em 19},~136.

\bibitem[Peskin and Galperin(2012)]{PeskinMGJCP12}
Peskin, U.; Galperin, M.
\newblock Coherently Controlled Molecular Junctions.
\newblock {\em J. Chem. Phys.} {\bf 2012}, {\em 136},~044107.

\bibitem[White \em{et~al.}(2013)White, Peskin, and
  Galperin]{WhitePeskinMGPRB13}
White, A.J.; Peskin, U.; Galperin, M.
\newblock Coherence in Charge and Energy Transfer in Molecular Junctions.
\newblock {\em Phys. Rev. B} {\bf 2013}, {\em 88},~205424.

\bibitem[Haug and Jauho(2008)]{HaugJauho_2008}
Haug, H.; Jauho, A.P.
\newblock {\em {Quantum {K}inetics in {T}ransport and {O}ptics of
  {S}emiconductors}}; Springer: Berlin/Heidelberg, Germany, 2008.

\bibitem[Aoki \em{et~al.}(2014)Aoki, Tsuji, Eckstein, Kollar, Oka, and
  Werner]{WernerRMP14}
Aoki, H.; Tsuji, N.; Eckstein, M.; Kollar, M.; Oka, T.; Werner, P.
\newblock Nonequilibrium Dynamical Mean-Field Theory and Its Applications.
\newblock {\em Rev. Mod. Phys.} {\bf 2014}, {\em 86},~779--837.

\bibitem[White \em{et~al.}(2014)White, Ochoa, and Galperin]{WhiteOchoaMGJPCC14}
White, A.J.; Ochoa, M.A.; Galperin, M.
\newblock Nonequilibrium Atomic Limit for Transport and Optical Response of
  Molecular Junctions.
\newblock {\em J. Phys. Chem. C} {\bf 2014}, {\em 118},~11159--11173.

\bibitem[Chen \em{et~al.}(2017)Chen, Ochoa, and Galperin]{ChenOchoaMGJCP17}
Chen, F.; Ochoa, M.A.; Galperin, M.
\newblock Nonequilibrium diagrammatic technique for {H}ubbard {G}reen
  functions.
\newblock {\em J. Chem. Phys.} {\bf 2017}, {\em 146},~092301.

\bibitem[Galperin(2017)]{MGChemSocRev17}
Galperin, M.
\newblock Photonics and spectroscopy in nanojunctions: A theoretical insight.
\newblock {\em Chem. Soc. Rev.} {\bf 2017},{\em 46}, 4000-4019.

\bibitem[Miwa \em{et~al.}(2017)Miwa, Chen, and Galperin]{MiwaChenMG17}
Miwa, K.; Chen, F.; Galperin, M.
\newblock Towards Noise Simulation in Interacting Nonequilibrium Systems
  Strongly Coupled to Baths.
\newblock {\em Sci. Rep.} {\bf 2017}, {\em 7},~9735.

\bibitem[Golizadeh-Mojarad and Datta(2007)]{DattaPRB07}
Golizadeh-Mojarad, R.; Datta, S.
\newblock Nonequilibrium Green's function based models for dephasing in quantum
  transport.
\newblock {\em Phys. Rev. B} {\bf 2007}, {\em 75},~081301.

\bibitem[Esposito \em{et~al.}(2015)Esposito, Ochoa, and
  Galperin]{EspOchoaMGPRB15}
Esposito, M.; Ochoa, M.A.; Galperin, M.
\newblock Nature of heat in strongly coupled open quantum systems.
\newblock {\em Phys. Rev. B} {\bf 2015}, {\em 92},~235440.

\bibitem[Ochoa \em{et~al.}(2016)Ochoa, Bruch, and Nitzan]{NitzanPRB16}
Ochoa, M.A.; Bruch, A.; Nitzan, A.
\newblock Energy distribution and local fluctuations in strongly coupled open
  quantum systems: The extended resonant level model.
\newblock {\em Phys. Rev. B} {\bf 2016}, {\em 94},~035420.

\bibitem[Gaspard(2015)]{GaspardNJP15}
Gaspard, P.
\newblock Scattering approach to the thermodynamics of quantum transport.
\newblock {\em New J. Phys.} {\bf 2015}, {\em 17},~045001.

\bibitem[Ness(2017)]{NessEntropy17}
Ness, H.
\newblock Nonequilibrium Thermodynamics and Steady State Density Matrix for
  Quantum Open Systems.
\newblock {\em Entropy} {\bf 2017}, {\em 19},~158.

\bibitem[Galperin and Nitzan(2006)]{GalperinNitzanJCP06}
Galperin, M.; Nitzan, A.
\newblock Optical Properties of Current Carrying Molecular Wires.
\newblock {\em J. Chem. Phys.} {\bf 2006}, {\em 124},~234709.

\bibitem[Scully(2010)]{ScullyPRL10}
Scully, M.O.
\newblock Quantum Photocell: Using Quantum Coherence to Reduce Radiative
  Recombination and Increase Efficiency.
\newblock {\em Phys. Rev. Lett.} {\bf 2010}, {\em 104},~207701.

\bibitem[Fetter and Walecka(1971)]{FetterWalecka_1971}
Fetter, A.L.; Walecka, J.D.
\newblock {\em {Quantum Theory of Many-Particle Systems}}; Dover Publications, Inc: Mineola, NY, USA,  2003.

\bibitem[Stefanucci and van Leeuwen(2013)]{StefanucciVanLeeuwen_2013}
Stefanucci, G.; van Leeuwen, R.
\newblock {\em Nonequilibrium Many-Body Theory of Quantum Systems. A Modern
  Introduction.}; Cambridge University Press: Cambridge, UK, 2013.

\bibitem[Jauho \em{et~al.}(1994)Jauho, Wingreen, and
  Meir]{JauhoWingreenMeirPRB94}
Jauho, A.P.; Wingreen, N.S.; Meir, Y.
\newblock Time-dependent transport in interacting and noninteracting
  resonant-tunneling systems.
\newblock {\em Phys. Rev. B} {\bf 1994}, {\em 50},~5528--5544.

\end{thebibliography}
\end{document}